
\magnification=1200
\hsize=6truein\vsize=8.5truein

\font\bigbf=cmbx10 scaled\magstep1


\def\frac#1/#2{\leavevmode\kern.1em
\raise.5ex\hbox{\the\scriptfont0 #1}\kern-.1em/\kern-.15em
\lower.25ex\hbox{\the\scriptfont0 #2}}
\def\sfrac#1/#2{\leavevmode\kern.1em
\raise.5ex\hbox{\the\scriptscriptfont0 #1}\kern-.1em/\kern-.15em
\lower.25ex\hbox{\the\scriptscriptfont0 #2}}

\def\gtorder{\mathrel{\raise.3ex\hbox{$>$}\mkern-14mu
             \lower0.6ex\hbox{$\sim$}}}
\def\ltorder{\mathrel{\raise.3ex\hbox{$<$}|mkern-14mu
             \lower0.6ex\hbox{\sim$}}}

\def\semidirprod{\rlap{\ss C}\raise1pt\hbox{$\mkern.75mu\times$}}

\def\for{\lower6pt\hbox{$\Big|$}}
\def\fish{\kern-.25em{\phantom{abcde}\over \phantom{abcde}}\kern-.25em}

\def\boxit#1{\vbox{\hrule\hbox{\vrule\kern3pt
        \vbox{\kern3pt#1\kern3pt}\kern3pt\vrule}\hrule}}
\def\dalemb#1#2{{\vbox{\hrule height .#2pt
        \hbox{\vrule width.#2pt height#1pt \kern#1pt
                \vrule width.#2pt}
        \hrule height.#2pt}}}


\def\noin{\noindent}

\def\al{\alpha}
\def\be{\beta}
\def\ga{\gamma}

\def\Ga{\Gamma}

\def\io{\iota}

\def\la{\lambda}

\def\si{\sigma}
\def\Si{\Sigma}

\def\cosec{{\rm cosec }}

\def\eg{{\it e.g. }}
\def\ie{{\it i.e. }}
\def\cf{{\it cf }}


\def\sup#1{${#1}$}

\def\reff#1{\smallskip\par\noindent{${#1}$}}

\def\Tr{{\rm Tr}}

\def\Z{\rlap{\rm Z}\mkern3mu{\rm Z}}
\def\R{{\rm\rlap I\mkern3mu R}}

\def\bra{\langle}
\def\ket{\rangle}
\def\ren{{\rm ren}}

\def\ref{\smallskip\par\noindent}

\def\sect{{\vskip 10truept\noindent}}

\def\3j#1#2#3#4#5#6{\left\lgroup\matrix{#1&#2&#3\cr#4&#5&#6\cr}
\right\rgroup}

\rightline {MUTP 91/30}
\vglue 1truein
\centerline{\bigbf POLYHEDRAL COSMIC STRINGS}
\vskip 15truept
\centerline{J.S.Dowker and Peter Chang}
\vskip 10truept
\centerline{\it Department of Theoretical Physics}
\centerline{\it The University,Manchester,England}
\vglue .75truein
\centerline{ABSTRACT}
\vskip 10truept
Quantum field theory is discussed in M\"obius corners using the method of
images. The vacuum average of the stress-energy tensor of a free field is
derived and is shown to be a simple sum of straight cosmic string
expressions. It does not seem possible to set up a spin-half theory easily.
\vskip 10truept
\rightline {May, 1992}
\vfill
\eject
\noindent{\bf 1 INTRODUCTION}\par\noindent

\noindent A straight, ideal cosmic string can be thought of as a conical
defect in space. A number of field theoretic calculations have been performed
in such a background which we do not wish to summarise here but simply refer
to a short review article\sup1. The attraction of this background is that it
is locally flat and the effect of the defect can be accommodated by changed
boundary (i.e. periodicity) conditions.

The defect is an angular one. The spatial metric can be written most naturally
in cylindrical coordinates
$$ds2=dz2+d\rho2+A2\rho2d\phi2\eqno(1)$$ where $\phi$ is a ``physical"
angle running from $0$ to $2\pi$. The deficit angle is $2\pi(1-A)$.

An object (a global monopole) with a defect in solid angle has been
discussed by Barriola and
Vilenkin\sup2 (see also Harari and Lousto\sup3, Mazzitelli and Lousto\sup4).
The metric is $$ds2=dr2+A2r2(d\theta2+\sin2\theta d\phi2).\eqno(2)$$
It is not locally flat, the curvature being proportional to $(1-A2)/r4$.
(Actually, such a metric was earlier suggested by Sokolov and Starobinsky\sup5
as an example of a metric with a conical singularity.)

In the present work we wish to discuss another situation with a solid
angle deficit that {\it is} locally flat. The price that must be paid is that
there are string singularities. Further,
the deficit can only take the values $4\pi(1-1/N)$ where $N$ is an integer. The
nontrivial values are thus much too large to be realistic. Nevertheless, the
geometry is interesting if only because its simplicity makes field theory
calculations easy and, in some ways, obvious. It is in this spirit that the
present calculation is presented.
\sect{\bf 2. THE BASIC CONSTRUCTION}.

\noindent The triangulation of a two-sphere obtained by its intersection with
the symmetry planes of an inscribed regular polyhedron is classic, if not
ancient. Consider such a triangulation applied to the constant $r$ sections
of the flat-space metric $$ds2=dr2+r2(d\theta'2+\sin2\theta'd\phi'2)
\eqno(3)$$ and join the triangle vertices to the origin $r=0$ to give a set
of ``triangular cones". We allow $r$ to extend to infinity.

If there are $2N$ triangles equivalent under the corresponding
extended point group, $\Ga'$, one obtains a division of the total solid
angle into $2N$ portions of $2\pi/N$, one for each cone.

One can now either treat such a cone as a physical region of $\R3$ and
proceed to
do (quantum) field theory in it with, say, Dirichlet or Neumann boundary
conditions or one can think of the cone as analogous to the segment
$0\le\phi'<\beta$ ($\beta=2\pi/N$) which, when its edges are identified,
yields the straight cosmic string (1) with $A=\beta/2\pi$.

For this latter interpretation it is necessary to be more careful regarding
the group action. Conventionally, the triangles are alternately shaded and
unshaded. The pure rotations of $\Ga'$, i.e. $\Gamma$, take shaded
to just shaded and unshaded to just unshaded regions. The remaining elements
interchange shaded and unshaded triangles and correspond to single
reflections (with
possibly a rotation as well). Thus, if we use the full extended group to
construct the Green function (of scalar field theory say) by the image method,
we will automatically obtain Dirichlet or Neumann boundary conditions
depending on how we combine the various contributions. Therefore this is
suited to our first interpretation of the triangular cone as a physical region.
The idea of the triangular cone, as a {\it trihedral kaleidoscope}, is
due to M\"obius\sup6. (See \eg Coxeter\sup{7,8}.)

To produce a {\it periodic} structure it is necessary to use the rotation
part, $\Gamma$, only  and to take the quadrilateral combination of a
triangle and one of its contiguous reflections, say, as the fundamental
domain on the two-sphere. This ``quadrilateral
cone" is the analogue of the segment in the straight cosmic string case
mentioned above. One expects the edges of the cone, which are just the
axes of symmetry of the corresponding regular solid, to be string-like.

If we adopt this attitude then the coordinates in (3), $(r,\theta',\phi')$,
are unphysical and it is necessary to find a coordinate transformation,
(analogous to $\phi=2\pi\phi'/\beta$ for the cosmic string) that takes the
quadrilateral cone to physical space i.e. onto an $\R3$. (Coordinates
on this $\R3$ will be our definition of `physical coordinates' although
nothing physical can, of course, depend on any particular choice.)

A possible coordinate transformation is
provided by the conformal transformation that takes a spherical triangle into
an upper half plane (and so takes the quadrilateral into the whole plane).
The calculations are, again, classic, the basic paper being that of
Schwarz\sup9. Of course, the field theory calculations are most easily done in
the metric (3).

A brief discussion of the wave equation in triangular cones was early given in
the book by Pockels\sup{10}, as noted by Laporte\sup{11} who further developed
the mode analysis. The reader will find in this paper a useful summary of
the situation. Keller\sup{12} includes the corners in his list of domains for
which the image method applies and there is mention of them, as `M\"obius
corners', in Terras and Swanson\sup{13} in connection with potential
problems. No doubt they occur elsewhere.

Developments of the mode problem from various points of
view are contained in Poole\sup{14}, Hodgkinson\sup{15}, Meyer\sup{16},
Stiefel\sup{17}, Altmann\sup{18}, Altmann and Bradley\sup{19}, Huber\sup{20}.

\sect {\bf 3. TRIANGULATIONS OF THE TWO-SPHERE.}

\noindent The classic account of the triangulations of S$2$ and their relation
to the finite dimensional subgroups of SO(3) (or SU(2)) is that of
Klein\sup{21} and its continuation in Klein and Fricke\sup{22}.
Discussions in English are less numerous but that by Forsyth\sup{23} is
useful and he also treats the conformal mapping by Schwarz triangle
(automorphic) functions. Further informative references are Ford\sup{24},
Cayley\sup{25},
Hurwitz and Courant\sup{26} and Caratheodory\sup{27}. A more recent source
of information and of further references is Coxeter\sup{7,8} . A
summary of Schwarz's theory can also be found in Darboux's treatise\sup{28}.

There are five basic cases corresponding to the cyclic, {\bf C}$_n$,
dihedral, {\bf D}$_n$, tetrahedral, {\bf T}, octahedral, {\bf O}, and
icosohedral, {\bf Y}, point groups, but the first two are more or less the
same and correspond to the straight cosmic string.

The angles of the fundamental triangles are written $\pi/\nu_i$, ($i=1,2,3)$,
where $(\nu_1,\nu_2,\nu_3)$ equals (2,3,3) for {\bf T}, (2,3,4) for {\bf O}
and (2,3,5) for {\bf Y}. The $\nu_i$ are related by $$\sum_i{1\over\nu_i}=
1+{2\over N}$$ and, if there are $n_i$ symmetry axes of the $\nu_i$-fold type,
$$\sum_i n_i(\nu_i-1)=N-1.$$ Further, $2(\sum_in_i-1)=N$.

In our situation, we start from the metric (3) where $\theta'$ and $\phi'$ are
standard angular coordinates on the unit sphere. The usual stereographic
projection onto the equatorial plane yields the Cayley-Klein parameter $w=u+iv$
which undergoes a linear fractional transformation (a homography or M\"obius
transformation) when the sphere is rotated. For a projection from the north
pole, $w=\cot(\theta'/2)\exp(i\phi')=(x'+iy')/(r-z')$ where $x',y',z'$ are
the cartesian coordinates of a
general point. $(r2=x'2+y'2+z'2)$. The metric (3) becomes
$$ds2=dr2+4r2{dwdw*\over(1+|w|2)2}.\eqno(4)$$

The conformal transformations, $w\rightarrow \zeta$ that map one $w$-triangle
onto the upper-half $\zeta$-plane are (e.g. Forsyth 1893)
$$1-\zeta={(w4+2iw2\sqrt3+1)3\over(w4-2iw2\sqrt3+1)3}\quad {\rm for}\quad
{\bf T}$$
$$\zeta={(w8+14w4+1)3\over(w{12}-33w8-33w4+1)2}\quad {\rm for}\quad {\bf
O}\eqno(5)$$
$$\zeta={(w{20}-22w{15}+494w{10}+228w5+1)3\over\big(w{30}+1+522w5(w{20
}-1)-10005w{10}(w{10}+1)\big)2}\quad{\rm for}\quad {\bf Y}.$$

The metric (4) is written equivalently
$$ds2=dr2+4r2\left|{dw\over d\zeta}\right|2{d\zeta d\zeta*\over(1+|w(
\zeta)|2)2}\eqno(6)$$
where $w$ is an algebraic function of $\zeta$ by (5). (6) explicitly shows the
singularities which conventionally have been chosen at $\zeta=0$, $\zeta=1$ and
$\zeta=\infty$. These positions can be altered by homographies applied to
$\zeta$ or to $w$ (Cayley\sup{25}).

We now think of $(r,\zeta,\zeta*)$ as coordinates in one to one correspondance
with the points of {\it physical} space. To emphasise this, angular variables,
$\theta$ and $\phi$ could be arbitrarily introduced by projecting the
$\zeta$-plane onto a unit
sphere by, say, $$\zeta=\cot(\theta/2)e{i\phi}.\eqno(7)$$ The entire
$\zeta$-plane is covered
by the ranges $0\le\theta<\pi$, $0\le\phi<2\pi$. The singularities lie at the
north and south poles and at one point on the equator of this $\zeta$-sphere.

There are three (cosmic) strings corresponding to the edges of the fundamental
quadrilateral (two vertices of which are identified). The ``strength" of a
string will be determined by its associated $\nu$ parameter i.e. by its
angular deficit. The strings meet at the origin, $r=0$.
\sect{\bf 4. THE CYCLIC GROUP C}$_n$ {\bf AND THE COORDINATE QUESTION}.

\noindent It is useful to discuss this elementary case as an example of our
general approach.

The triangulation of the unit sphere is a division into $n$
lunes running from pole to pole, which are the only fixed points. The $z'$ axis
is taken as the (polar) rotation axis and the stereographic projection onto the
$(x',y')$-plane is a set of $n$ wedges of angle $2\pi/n$. The standard
conformal
transformation is $$wn=\zeta$$ which opens out each wedge into the whole
$\zeta$-plane. The singularity (the string) in $\R3$ lies on the rotation
axis.

The metric (6) will be a complicated function of the new colatitude,
$\theta$, (in (7)). In order to regain the ``natural" metric, (1), a coordinate
transformation from $\theta$ back to $\theta'$ using
$(\cot(\theta'/2))n=\cot(\theta/2)$ is necessary and then $\rho$ and $z$
can be introduced conventionally. The result of all this is just to
rescale the azimuth $\phi'$ to $\phi$ and to leave the colatitude, $\theta'$,
alone. This is what one does immediately, of course, but the general case is
not so clear and the conformal transformation seems to be the only
systematic way of obtaining ``physical" coordinates.

The remaining problem in the general case is the interpretation of the
coordinates. That is to say, what is the physical significance of the
singularities ? In the cyclic case one can talk about lensing effects, for
example. Is there anything similar for the other cases? Is it possible to
produce a source for the singularities and does anything special happen at
the origin where the three strings meet ?
\sect{\bf 5. QUANTUM FIELD THEORY}.

\noindent We abandon the question of physical significance and
return to the metric (3). The point $(\theta',\phi')$ on the unit sphere
is denoted by $q$ and the elements of $\Ga'$ by $\gamma$. The action
of $\Ga'$ on the unit sphere can be extended in the obvious way to $\R3$
by $\gamma{\bf r}=\gamma(r,q)=(r,\gamma q)$. The triangular cone, ${\cal C}'$,
can be taken as the fundamental domain of $\Ga'$ acting on all of
three-space,$${\cal C}'=\R3/\Ga'\approx\R+\times{\rm
S}2/\Ga'.$$ The quadrilateral fundamental domain is ${\cal C}=\R3/\Gamma$.

In the case of free scalar field
theory, or quantum mechanics, the Green function, or propagator, in the
corner will be given by an image sum of standard, Minkowski Green
functions, or of standard Euclidean propagators. Typically,
$$G(r,q,t;r',q',t')=\sum_{\gamma\in\Ga'}a(\gamma)G_0(r,q,t;r',\gamma q',t')
.\eqno(8)$$ The phase factors $a(\gamma)$, in the simplest cases, are either
all equal to unity (giving Neumann boundary conditions) or equal to one
when $\gamma$ is a rotation and minus one otherwise (giving Dirichlet
conditions).

As mentioned earlier, to produce periodic conditions the summation in (8)
should be restricted to the group, $\Gamma$, of pure rotations. The phase
factors are then all unity. (For complex fields, we might allow the $a(\gamma)$
to be more general phases. This will be discussed later.)

The coordinates $(r,q)$ on the left hand side of (8) are, of course, restricted
to the fundamental domains, ${\cal C}'$ or ${\cal C}$ as the case may be.

It is often considered interesting to calculate vacuum averages of various
operators, particularly $\bra\phi2\ket$ and $\bra T_{\mu\nu}\ket$, as evidence
of a Casimir effect. For example,
$$\bra\phi2\ket=-i\lim_{x'\rightarrow x}G(x;x')$$where $x$ stands for
$(r,q,t)$. Equation (8) shows that this will diverge and we remove the
offending term, $\gamma={\rm id}$, from the sum as our renormalisation.
Then $$\bra\phi2\ket_\ren=-i\sum_{\gamma\ne{\rm id}}a(\gamma)G_0(r,q,t;r,
\gamma q,t).\eqno(9)$$ The unit element of a group is denoted by ${\rm id}$
or, sometimes, by $E$.

For a massless field $G_0(x,x')=-(i/4\pi2)(1/\la2)$, as a distribution,
where $\la2=(t-t')2-|{\bf r-r'}|2$. Therefore
$$\bra\phi2({\bf r})\ket_\ren=
{1\over 4\pi2r2}\sum_{\gamma\ne{\rm id}}a(\gamma){1\over d(q,\gamma q)2}.
\eqno(10)$$
where $d(q,q')$ is the Euclidean distance between $(1,q)$ and $(1,q')$.

For complex fields the equation corresponding to (10) is
$$\bra|\phi|2({\bf r})\ket_\ren={1\over 2\pi2r2}\sum_{\gamma\ne{\rm id}}
{\rm Re}\big(a(\gamma)\big){1\over d(q,\gamma q)2}.\eqno(11)$$

Restricting attention for the moment to the periodic case, the sum over
$\Gamma$ in (10) can be recast into a sum over the fixed points (or axes).
Denote a typical symmetry axis, (= two fixed points or vertices), by
${\bf k}$. Let the associated parameter be $\nu_{\bf k}$ (one of the $\nu_i$)
and the corresponding generator $A_{\bf k}$. Simple geometry gives
$$rd(q,A_{\bf k}m q)=2\sin(m\pi/\nu_{\bf k})\rho({\bf r},{\bf k}),$$
where $\rho({\bf r},{\bf k})$ is the perpendicular Euclidean distance from
${\bf r}$ to the axis ${\bf k}$,
so that (10) becomes (we drop the `ren' suffix)$$\bra\phi2({\bf r})\ket=
{1\over 16\pi2}\sum_{{\bf k}}\left[\sum_{m=1}{\nu_{\bf k}-1}\cosec2
\left({m\pi\over\nu_{\bf k}}\right)\right]{1\over \rho({\bf r},{\bf k})2}$$
$$={1\over 48\pi2}\sum_{{\bf k}}{\nu_{\bf k}2-1\over \rho({\bf r},{\bf k})2}
=\sum_{{\bf k}}\bra\phi2({\bf r})\ket_{\nu_{\bf k},{\bf k}}.
\eqno(12)$$

This is just the sum of the $\bra\phi2({\bf r})\ket$'s for each of
the singularity axes taken separately.

A similar result holds for other vacuum averages. This can be shown quite
generally, and easily, by writing the sum over $\Ga$ in (8) as one over
the axes ${\bf k}$ and the associated cyclic operations,

$$G({\bf r},t;{\bf r}',t')=G_0({\bf r},t;{\bf r}',t')+\sum_{\bf k}\sum_{m=1}{
\nu_{\bf k}-1} G_0({\bf r},t;A_{\bf k}m {\bf r}',t')$$
\quad\qquad$$=G_0({\bf r},t;{\bf r}',t')+\sum_{\bf k}G_{\bf k}({\bf r},t;{\bf
r}',t').\eqno(13)$$
$G_0+G_{\bf k}$ is the Green function for a singularity axis, ${\bf k}$.
Invariance under the general rotation $R$ says that
$$G_0(R{\bf r},t;A_{R\bf k}R{\bf r}',t')=G_0({\bf r},t;A_{\bf k}
{\bf r}',t')$$where $A_{R\bf k}=RA_{\bf k}R{-1}$ is a generator about
the axis $R{\bf k}$.

All renormalised vacuum averages then have a similar structure. That of
the stress-energy tensor
is given by $$\bra T_{\al\be}({\bf r})\ket=\sum_{\bf k}\bra T_{\al\be}
({\bf r})\ket_{\nu_{\bf k},\bf k}\eqno(14)$$
where $\bra T_{\al\be}({\bf r})\ket_{\nu,\bf k}$ is the vacuum
average around a cosmic string of angle $2\pi/\nu$ along the axis ${\bf k}$.
This expression allows for a simple numerical evaluation since the
individual vacuum averages are already known. We must note,
of course, that the tensor indices in (14) refer to coordinates on the $\R4$
covering space so that the standard result for a single, straight string along
the $z$-axis must be transformed to strictly Minkowskian coordinates in $\R4$
and then rotated from ${\bf z}$ to ${\bf k}$. Performing these operations,
the general form of $\bra T{\al\be}({\bf r})\ket$ for a string lying along
the axis ${\bf k}$ is found to be,
$$\bra T{\al\be}({\bf r})\ket_{\nu,\bf k}={C_\nu\over16\pi2\rho6}[
\rho2\eta{\al\be}+4v\al v\be],\eqno(15)$$where $C_\nu$ is a constant
depending on the deficit angle and the vector $v\al=(0,{\bf k}\wedge{\bf r})$.
This expression is for conformally-invariant, massless fields.

In Fig.1 we show a contour plot for a periodic
conformal scalar. We have chosen the octahedral group for illustrative
purposes and have pictured $\bra T{00}({\bf r})\ket$ for $r=1$
over the equatorial plane of the stereographic projection of the triangulation,
which is also shown. Fig.2 shows a relief plot of the same quantity.

The plots for the Dirichlet and Neumann boundary conditions are generally
similar. We give in Fig.3 a contour plot of the tetrahedron Dirichlet case.
\sect{\bf 6. PHASE FACTORS}.

\noindent In quantum mechanics, or in the theory of a complex field, one is
allowed to choose the phase factors in (8) to be a unitary representation of
the ramified covering group of ${\cal C}'$(${\cal C}$), which is $\Ga'$($
\Gamma$). The character tables indicate nontrivial one-dimensional
representations in the {\bf T} and {\bf O} cases. To establish some other
notation, we check these facts in a standard way.
Consider the presentation $A_1{\nu_1}=A_2{\nu_2}=A_3{\nu_3}=A_1A_2A_3=
E$ where $A_i$ generates rotations through $2\pi/\nu_i$ about the
vertex of angle $\pi/\nu_i$. The abelian nature of the representation
$\{a(\gamma)\}$ implies that $$a(\gamma)=e{2\pi i{\bf p.q}}\eqno(16)$$
where $q_i=s_i/\nu_i$ and $p_i$ is the number of times the generator $A_i$
occurs in the word presentation of $\gamma$. The final relation translates
into $\sum q_i\in\Z$ with $0\le q_i<1$.

Analysing the cases yields the complex representations ${\bf s}=(0,2,1)$ or
$(0,1,2)$ for {\bf T} and the real one $(1,0,2)$ for {\bf O}.
 (Equivalent statements are $H_1(\R3/{\bf T};\Z)\approx\Z_3$ and
$H_1(\R3/{\bf O};\Z)\approx\Z_2$. Also $H_1(\R3/{\bf Y};\Z)\approx{\bf 0}$.)

For {\bf T}, one interpretation of (0,2,1), in the covering space picture,
would be of 4 strings of two units of flux running from the origin
to the four vertices and four strings of unit flux to the four face centres
(which are the vertices of the counter-tetrahedron). Such a flux
distribution has no effect on the quantum mechanics on the covering space
so that the Green function, or propagator, $G_0$, in (8) is unaffected.

Nothing physical is altered by
adding three flux units along any of the $\nu=3$ rotation axes and we have
the equivalent interpretation as four U(1) Aharonov-Bohm flux tubes along
these axes all carrying either one unit or two units of flux corresponding,
respectively, to the two cases (0,2,1) and (0,1,2).

Results analogous to (12), (13) and (14) hold with the appropriate
modifications of the individual cosmic string contributions.

Figs.4 shows a relief plot of
$\bra T{00}({\bf r})\ket$ for the (0,2,1) tetrahedron case.

The above discussion can be generalised by taking the field to belong to a
representation of some non-abelian internal symmetry group, G. Then the
$a(\ga)$ will be elements of Hom$_{\rm G}(\Ga,$G) and we can use our knowledge
of the representations of $\Ga$ to construct this homomorphism. Since the
vacuum averages involve a trace over the internal indices,
expressions like (10) will contain the G-character, $\chi(\ga)$.

A simple example would be $\phi$ in the fundamental representation of
G=U(2) and ${a(\ga)}$ the $2\times2$ irreducible representation of
${\bf O}$ usually denoted by E. G has been chosen to be
U(2) rather than SU(2) to allow the use of the irrep E, which is not
unimodular.

A glance at the character tables shows that
certain classes of ${\bf O}$ will disappear from sums such as (10). The result
is the same as that for the unimodular representation, E$\oplus$E$*$, in the
{\bf T} case. Thus, so far as certain vacuum averages are concerned, the
gravitational effect of a singular string can be removed by a suitably
contrived arrangement of (non-abelian) internal symmetry fluxes.
\sect{7. \bf OTHER FIELDS}

\noin For simplicity, we will discuss the periodic case first, and indicate the
modifications needed for the extended group $\Ga'$ later.

\sect{\bf The electromagnetic field}.

\noin For brevity we use the $\phi={\bf H}-i{\bf E}$ formalism employed
elsewhere in a similar context. Using Cartesian axes in the covering $\R3$
space, the relevant vector Green function $G_0$ is a $3\times3$ matrix and
rotational invariance now reads
$$G_0(R{\bf r},t;A_{R\bf k}R{\bf r}',t')=D(R)G_0({\bf r},t;A_{\bf k}
{\bf r}',t')D(R{-1})\eqno(17)$$where $D(R)$ is a spin-one representation of
SO(3).

It is possible to extend this formalism to the spin-$j$ case and we will
imagine this to be done simply by taking the matrices to be $(2j+1)\times
(2j+1)$ ones.

The projected Green function on T$\times\R3/\Ga$ is
$$G({\bf r},t;{\bf r}',t')=\sum_{\gamma\in\Ga}a(\ga)G_0({\bf r},t;\ga{\bf r}'
,t')D(\ga)=\sum_{\gamma\in\Ga}a(\ga)D(\ga)G_0(\ga{-1}{\bf r},t;{\bf r}',
t')\eqno(18)$$where the $D(\ga)$ factor arises from a rotation between a
local dreibein system (invariant under $\Ga$) and the globally Cartesian
one.(\cf Banach and Dowker\sup{29} equation (A15).) The U(1) factors, $a(\ga)$,
have been retained for generality.

The summation can again be written over the singularity axes, as in (13),
but where $G_{\bf k}$ is this time given by
$$G_{\bf k}({\bf r},t;{\bf r}',t')=\sum_{m=1}{\nu_{\bf k}-1}a(A_{\bf k}m
)G_0({\bf r},t;A_{\bf k}m{\bf r}',t')D(A_{\bf k}m).\eqno(19)$$

$G_{\bf k}$ is the spin-$j$ Green function for a singularity axis, ${\bf k}$,
in Cartesian coordinates and with respect to a globally Cartesian dreibein
system.

In the electromagnetic spin-one case, the vacuum average of the energy density
is proportional to $\Tr G$ and we see from (18) and (19) that it has the
same form as in (14). The same statement, \ie (14) with (15), holds for the
complete stress-energy tensor.
\sect{\bf The spin-half field}

\noin There appears to be an obstruction to setting up a spinor field on a
fundamental domain, ${\cal C}$. The reason is that the image method for
deriving the Green function around a single, straight cosmic string, \ie
in the cyclic $\Z_\nu$
case, does not work if $\nu$ is even. This problem can be circumvented for
the straight string by artificially introducing a U(1) flux through the
string so as to give an extra minus sign when the string is encircled.
In the ${\bf T}$, ${\bf O}$ and ${\bf Y}$ cases this is not possible
because of the relation, $A_1A_2A_3=E$, between the generators.

It is possible to put spinors around the general straight cosmic string but
the Green function can only be written as an image sum
when the angle is $2\pi/\nu$ with $\nu$ odd. It is not clear whether spinors
can be set up easily on ${\cal C}$ without using images.
\sect{\bf8. THE EXTENDED GROUPS}.

\noin The extended groups are the complete symmetry groups of the regular
solids. In the Sch\"onflies notation, they are denoted by ${\bf T}_d$,
${\bf O}_h$ and ${\bf Y}_h$.

They can be generated by the planes of symmetry, in particular by
reflections in three (concurrent) planes (\eg Coxeter\sup8 \S 5.4) and,
as mentioned before, the elements fall into two sets depending on whether
they contain an even or an odd number of reflections. Even, or proper, elements
can be written as $\ga\in\Ga$ and odd, or improper, elements as $\ga\si$,
where $\si$ is a reflection in a symmetry plane of the regular solid.

The image sum (18) then becomes
$$G({\bf r},t;{\bf r'},t')=\sum_{\gamma\in\Ga}a(\ga)\bigg(G_0({\bf r},t;
\ga{\bf r}',t')+a(\si)G_0({\bf r},t;\ga\si{\bf r}',t')
\Si(\si)\bigg)D(\ga),\eqno(20)$$where $\Si(\si)$ is the action on the field
induced by $\si$.

Actually, because a reflection mixes left and right, it is
necessary to extend the $(2j+1)\times(2j+1)$ formalism to a
$2(2j+1)\times2(2j+1)$ Dirac one. For neutral fields we can write $\psi=
\phi\oplus C\phi*$ where $C$ is the charge conjugation matrix.
The $D(\ga)$ in (20) should thus be taken as a direct sum representation,
$$D(\ga)=D(\ga)\oplus D(\ga)=D(\ga)\otimes{\bf 1}.$$Choosing
$\phi={\bf H}-i{\bf E}$ corresponds to setting $C$ equal to the unit matrix.

The reflection, $\si_{\bf t}$, in the plane with normal ${\bf t}$,
can be written as a rotation through $\pi$ about the
axis ${\bf t}$ combined with the parity inversion $\io:{\bf r}\rightarrow
-{\bf r}$, \ie $\si_{\bf t}=R_{\bf t}(\pi)\io$. For arbitrary spin, we have
the reflective action $$\big(R_{\bf t}(\pi)P\psi\big)({\bf r},t)=\Si(\si_
{\bf t})\psi(\si_{\bf t}{\bf r},t)$$where $P$ is the usual parity action
$$\big(P\psi\big)({\bf r},t)={\bf 1}\otimes\si1\psi(-{\bf r},t),$$
$\si1$ being the standard Pauli matrix.
Thus the representation $\Si(\si_{\bf t})$ is given by
$$\Si(\si_{\bf t})=D\big(R_{\bf t}(\pi)\big)\otimes\si1.$$
Explicitly for spin-one, in the Cartesian basis ($C={\bf 1}$),
$$D\big(R_{\bf t}(\pi)\big)=2{\bf t}\otimes{\bf t}-{\bf 1}.$$

The action on $\phi={\bf H}-i{\bf E}$ corresponds to complex conjugation
together with a reversal in sign of the parallel component while the sign
of the normal part is unchanged. Thus $a(\si)$ should be set equal to $-1$
in order to give the correct (perfect) boundary conditions in ${\cal C}'$
for the electromagnetic field.

The reflection term in $G({\bf r},t;{\bf r'},t')$ disappears trivially from
$\Tr G$. Therefore the vacuum average of the electromagnetic stress tensor
is the same in the periodic and reflective cases.

For the cyclic wedge this
agrees with known, and old results (\eg Candelas and Deutsch\sup{30}). In this
simple geometry the result also holds for the conformally coupled scalar
field because the contribution to the average coming from single reflection
terms vanishes. This also agrees with old results. There is, however, a
difference in the polyhedral cases for the scalar field.

For the octahedron and icosahedron, the extended groups can be composed as
$\Ga'=\Ga\cup\Ga\io$. The inversion $\io$ can be expressed in terms of the
generating reflections,
$\si_1$, $\si_2$, $\si_3$, in the sides of a fundamental triangle, as
$$\io=(\si_1\si_2\si_3){h/2},$$where $h$ is the Coxeter-Killing number
connected with the order of the group  by $|\Ga'|=h(h+2)$.

In general,
$h$ is the period of the product of the reflection generators of a finite
reflection group. When $h/2$ is even, $(\si_1\si_2\si_3){h/2}$ is a
half-turn. The generators $A_i$ introduced earlier are related to
the $\si_i$ by $A_1=\si_3\si_2$ {\it etc}.

In the cases $O_h$ and $Y_h$, (20) can be replaced by
$$G({\bf r},t;{\bf r'},t')=\sum_{\gamma\in\Ga}a(\ga)\bigg[G_0({\bf r},t;
\ga{\bf r}',t')D(\ga)\otimes{\bf 1}\pm G_0({\bf r},t;-\ga{\bf r}',t')
D(\ga)\otimes{\si1}\bigg].\eqno(21)$$

Incidentally, a light ray sent into a tetrahedral corner will emerge, after
six reflections, in a different direction whereas, in the octahedral and
icosahedral cases, it will come out simply reversed (and translated).

\sect{\bf9. CONCLUSION AND COMMENTS}.

\noin The method of images has been used to set up non-interacting field
theories in M\"obius corners. The physical system corresponds to a
combination of three concurrent ideal cosmic strings. Some Casimir effects
have been calculated and, although somewhat academic, the expressions are, we
feel, sufficiently attractive to warrant exposure.

We still have to address the questions raised in section 4, particularly
the production of a source for the metric.

In the cyclic case, {\bf C}$_n$, one can introduce $m$-fold coverings of the
sphere with a lune of angle $2\pi m/n$ as fundamental domain, ($m\in\Z)$.
Letting $m$ tend to infinity one gets an infinitely sheeted covering (cf
Sommerfeld\sup{31}) and a wedge of arbitrary angle can be treated.

For the other cases it is only possible to find a finite number of finitely
sheeted coverings (Schwarz\sup9). It is not clear whether this means that
one cannot find a way of treating an arbitrary solid angle deficit. The
corresponding conformal transformation in terms of hypergeometric functions
is standard, but the problem is the construction of the Green function on
the multi-sheeted Riemann surface. Unless this can be done, and this is an
open problem, there is little point in analysing the possible
observational significance of the metric (6), except as a mathematical
exercise. Serebryanyi\sup{32} gives some interesting generalities on the
covering space method.

At a calculational level the group theoretical analysis of the mode problem
and the Clebsch-Gordan series is interesting and has been looked at in some
depth by the chemists e.g. Damhus et al \sup{33}.
\vfill\eject
\centerline{FIGURE CAPTIONS}
\sect Fig.1. Contour plot of the vacuum average of the energy density of a
massless, conformally coupled scalar field for periodic boundary conditions in
an octahedral M\"obius corner. The section is for constant radius $r=1$ and
the horizontal axes are those of the equatorial plane of the stereographic
projection which is also depicted.
\sect Fig.2. Relief plot of Fig.1.
\sect Fig.3. Contour plot of the vacuum average of the energy density of a
massless, conformally coupled scalar for Dirichlet boundary conditions in
a tetrahedral M\"obius corner.
\sect Fig.4. Relief plot of the vacuum average of the energy density of a
massless, conformally coupled complex scalar with twisted boundary
conditions in a tetrahedral corner.
\vfill\eject

\sect{\bf REFERENCES}
\reff1 Dowker,J.S. 1990 ``Field theory around conical defects" in ``The
formation and evolution of cosmic strings" ed. by S.W.Hawking, G.W.Gibbons and
T.Vachaspati (Cambridge University Press, Cambridge).
\reff2 Barriola,M. and Vilenkin,A. 1989 Phys.Rev.Lett.{\bf 63} 341.
\reff3 Harari,D. and Lousto,C. 1989 Phys.Rev.{\bf D}.
\reff4 Mazzitelli,F.D. and Lousto,C. 1991 Phys.Rev.{\bf D43} 468.
\reff5 Sokolov,D.D. and Starobinsky,A.A. 1977 Dokl.Akad.Nauk. {\bf 234} 1043.
\reff6 M\"obius,A.F. 1849 Ver.K\"on.S\"ach.Ges.Wiss. {\bf 1} 65. Reprinted
1852 J.f.Math.(Crelle) {\bf 43} 365.

\reff7 Coxeter,H.S.M. 1974 ``Regular complex polytopes" (C.U.P.,Cambridge).
\reff8 Coxeter,H.S.M. 1948 ``Regular polytopes" (Methuen,London).
\reff9 Schwarz,H.A. 1873 J.f.Math.(Crelle){\bf 75},292.
\reff{10} Pockels,F. 1891 ``Uber die Differentialgleichung $\Delta u+k2u=0"$
(Teubner,Leipzig).
\reff{11} Laporte,O. 1948 Zeit.f.Naturf.{\bf 3a},447.
\reff{12} Keller,J.B. 1953 Comm.Pure.Applied Math. {\bf 6} 505.
\reff{13} Terras,A. and Swanson,R. 1980 J.Math.Phys. {\bf 21} 2140.
\reff{14} Poole, E.C.G. 1932 Proc.Lond.Math.Soc.{\bf 33},435.
\reff{15} Hodgkinson,J. 1935 J.Lond.Math.Soc.{\bf 10},221.
\reff{16} Meyer,B. 1954 Can.J.Math.{\bf 6} 135.
\reff{17} Stiefel,E. 1952, J.Res.N.B.S.{\bf 48},424.
\reff{18} Altmann,S.L. 1957 Proc.Camb.Phil.Soc.{\bf 53} 343.
\reff{19} Altmann,S.L. and Bradley,C.J. 1963 Trans.Roy.Soc.{\bf 255} 199.
\reff{20} Huber,H. 1970 Comm.Math.Helv.{\bf 45} 494.
\reff{21} Klein,F. 1884 ``Vorlesungen \"uber das Ikosaeder" (Teubner,Leipzig).
2nd.Edition \break Translated by G.C.Morrice as ``Lectures on the icosahedron"
(Methuen, London, 1913).
\reff{22} Klein,F. and Fricke,R. 1890 ``Vorlesungen \"uber die Theorie der
elliptischen Modulfunctionen" (Teubner,Leipzig).
\reff{23} Forsyth,A.R. 1893 ``Functions of a complex variable." (C.U.P.,
Cambridge).
\reff{24} Ford,L.R. 1929 ``Automorphic functions" (McGraw-Hill,New York).
\reff{25} Cayley,A. 1879 Trans.Camb.Phil.Soc. {\bf 13},5.
\reff{26} Hurwitz,A. and Courant,R. 1925 ``Funktionentheorie" (Springer,
Berlin).
\reff{27} Caratheodory,C. 1953 ``Theory of functions" vol.2 (Chelsea,New York).

\reff{28} Darboux,G. 1887 ``Th\'eorie g\'en\'erale des surfaces" vol.1. p.170.
\reff{29} Banach,R. and Dowker,J.S. 1979 J.Phys. A{\bf 12} 2545.
\reff{30} Deutsch, D. and Candelas,P. 1979 Phys.Rev. {\bf D20} 3063.
\reff{31} Sommerfeld, A. 1897 Proc.Lond.Math.Soc.{\bf 28} 395
\reff{32} Serebryanyi,E.M.1982 Theor.Math.Phys.{\bf 52} 651.
\reff{33} Damhus,T.,Harnung,S.E. and Schaffer,C.E. 1984 Theor.Chim.Acta.{\bf
65}
317.
\end